%% file: main.tex
\pdfoutput=1
\documentclass[acmlarge]{acmart}
\usepackage{amsmath}
\usepackage{graphicx}
\usepackage{xcolor}
\usepackage{soul}
\usepackage{amsfonts}
\usepackage[noend]{algorithmic}
\usepackage[noend,ruled,linesnumbered,longend,lined]{algorithm2e}
\usepackage{amsmath}
\usepackage{float}
\usepackage[caption=false]{subfig}
\usepackage{diagbox}

\begin{document}

\title{An Ensemble Learning Approach for Exercise Detection in Type~1 Diabetes Patients}


\author{Ke Ma}
\affiliation{%
 \institution{Stony Brook University}
  \streetaddress{100 Nicolls Rd}
  \city{Stony Brook}
    \state{NY}
  \country{USA}}
 \email{make79712@gmail.com}

\author{Hongkai Chen}
\affiliation{%
  \institution{Stony Brook University}
  \streetaddress{100 Nicolls Rd}
  \city{Stony Brook}
    \state{NY}
  \country{USA}}
\email{hongkai.chen@stonybrook.edu}
\orcid{0000-0001-7206-6584}

\author{Shan Lin}
\affiliation{%
  \institution{Stony Brook University}
  \streetaddress{100 Nicolls Rd}
  \city{Stony Brook}
  \state{NY}
  \country{USA}
  \postcode{11790}}
\email{shan.x.lin@stonybrook.edu}
\orcid{0000-0001-6362-2972}

\input{abstract}

\maketitle

\input{introduction}
\input{related_work}
\input{machine_learning_model}
\input{glucose_model}
\input{combine_both_models}

\input{experiment_setup}

\input{experiment_result}
\input{future_work}
\input{Conclusion}

\bibliographystyle{ACM-Reference-Format}

\bibliography{mybibliography}

\end{document}

%% file: abstract.tex
\begin{abstract}

Type 1 diabetes is a serious disease in which individuals are unable to regulate their blood glucose levels, leading to various medical complications. Artificial pancreas (AP) systems have been developed as a solution for type 1 diabetic patients to mimic the behavior of the pancreas and regulate blood glucose levels. However, current AP systems lack detection capabilities for exercise-induced glucose intake, which can last up to 4 to 8 hours. This incapability can lead to hypoglycemia, which if left untreated, could have serious consequences, including death.
Existing exercise detection methods are either limited to single sensor data or use inaccurate models for exercise detection, making them less effective in practice. 
In this work, we propose an ensemble learning framework that combines a data-driven physiological model and a Siamese network to leverage multiple physiological signal streams for exercise detection with high accuracy.
To evaluate the effectiveness of our proposed approach, we utilized a public dataset with  12 diabetic patients collected from an 8-week clinical trial. Our approach achieves a true positive rate for exercise detection of $86.4\%$ and a true negative rate of $99.1\%$, outperforming state-of-the-art solutions.
\end{abstract}


%% file: introduction.tex
\section{Introduction}

Type 1 Diabetes Mellitus (T1DM) is a chronic autoimmune disease that affects the human pancreas' ability to produce sufficient insulin to regulate blood glucose (BG) levels. This results in hyperglycemia (high BG), which can lead to various health complications such as cardiovascular disease, kidney damage, and blindness~\cite{chen2019committed}. Insulin infusion from external sources is necessary for T1DM patients~\cite{hernandez2008extension}. 
The \textit{artificial pancreas} (AP) is a system for automatically delivering insulin for T1DM patients~\cite{paoletti2019data}.
The AP consists of an insulin infusion pump and a subcutaneous continuous glucose monitor (CGM) for sensing BG levels.
However, the lack of physiological counter-regulatory response to exercise-induced decreases in BG levels can result in potentially dangerous hypoglycemia if insulin infusion is not adjusted accordingly~\cite{dasanayake2015early}.

Exercise has been shown to increase insulin sensitivity and glucose uptake in the body, which can lead to a sharp decrease in BG levels~\cite{jacobs2015incorporating}. In healthy individuals, this decrease is compensated by a drop in insulin levels. However, T1DM patients cannot adjust insulin levels rapidly in response to exercise, leading to excess insulin promoting amplified glucose uptake in skeletal muscles and suppressing endogenous glucose production. This can increase the probability of a hypoglycemic event, making it crucial to regulate blood glucose for diabetes patients by detecting exercise automatically.
While meal perturbations are satisfactorily taken into account in existing models, exercise detection models for diabetes patients are lacking. Current models use simulation data and do not consider time delays in detection~\cite{ramkissoon2019detection,dasanayake2015early,zakeri2008application}. 

T1DM patients face the challenge of regulating their blood glucose levels during exercise, which can lead to health complications if not managed properly.
To overcome this challenge, we propose an ensemble learning-based approach that combines a data-driven model and a Siamese network to detect exercise events in T1DM patients with high accuracy. The Siamese network-based detector takes sequences of physiological signals, including galvanic skin response (GSR), heart rate, steps count, and skin temperature, and detects the difference between non-exercise and exercise samples. Additionally, we propose a new model that considers the effect of exercise and meal intake on glucose levels and uses BG readings to estimate exercise-related variables. We estimate the known parameters of the model using the OhioT1DM dataset~\cite{marling2020ohiot1dm}, which contains 8 weeks of physiological data for 12 T1DM patients. By automating exercise detection and adjusting dosing based on the level and type of exercise, our approach has the potential to mitigate the risk of exercise-induced hypoglycemia in T1DM patients. We compare our approach with four baseline approaches and achieve $86.4\%$ detection accuracy and $99.1\%$ true negative rate, outperforming state-of-the-art approaches. Our results demonstrate the effectiveness of our approach in improving the detection of exercise events in T1DM patients.

%% file: related_work.tex
\section{related work}
Several studies have attempted to detect exercise in T1DM patients using different physiological signals.
For instance, Dasanayake et al.~\cite{breton2008physical} utilize PCA and T statics techniques to analyze simulated heart rate data and acceleration data. However, these models overlook glucose data, a significant indicator of exercise.
Similarly, Breton et al.~\cite{breton2008physical} only consider heart rate to quantify the effect of exercise.
In contrast, Zakeri et al.~\cite{zakeri2008application} estimate energy expenditure by assuming a linear relationship between heart rate, acceleration data, and energy expenditure. However, this assumption may not always hold true, e.g., energy expenditure is often considered nonlinear to heart rate~\cite{billat2006nonlinear}.
In Ramkissoon et al.~\cite{ramkissoon2019detection}, the authors add a parameter to the Bergman minimal model~\cite{Bergman81physiologicevaluation} to describe plasma glucose variations. 
However, this parameter is a proxy signal to the blood glucose variation due to different disturbances, such as missing meals or overdoses of insulin.
Other studies such as Dalla Man et al.~\cite{dalla2009physical} and Hernández-Ordoñez et al.~\cite{hernandez2008extension}  proposed exercise-glucose models that consider the effect of exercise on blood glucose levels. However, these models are limited in their consideration of different levels of exercise or the insulin action is still proportional to heart rate even when patients are not exercising. In contrast, our proposed approach considers multiple signals from wearable devices, including step counts, heart rate, GSR, and skin temperature, to detect exercise events and estimate exercise-related variables.

Ensemble learning has gained increasing attention in recent years for its ability to combine multiple heterogeneous models to achieve better performance than individual models. It has been widely used in various fields, including computer vision, signal processing, and speech recognition~\cite{huang2019magtrack,sagi2018ensemble}. 
The basic idea of ensemble learning is to train multiple models with different features, algorithms, or hyper-parameters and then combine their predictions to improve accuracy and robustness. 

%% file: machine_learning_model.tex
\section{System Overview}\label{sec:sys}

In this section, we provide an overview of our proposed approach for exercise detection in T1DM patients using an ensemble learning method. 
Our approach combines machine learning models and data-driven physiological models to improve the accuracy and overall performance of exercise detection.
To detect exercise events, we use a Siamese network-based detector that takes sequences of physiological signals, including GSR, heart rate, step counts, and skin temperature. The Siamese network is trained to detect the difference between exercise and non-exercise samples. We also proposed a new model that considers the effect of exercise and meal intake on glucose and uses BG readings to estimate exercise-related variables to detect exercise events.

\subsection{Machine Learning Models}\label{subsec:ml_model}

In this section, we describe our use of a Siamese network for exercise detection. 
The Siamese network takes sequences of physiological signals, including GSR, heart rate, steps count, and skin temperature, as inputs. 
Using a non-exercise sample as one of its inputs, the Siamese network detects the difference between that and the other input, and if the difference is large, it determines the input as an exercise event.

\subsubsection{Siamese Networks}
\begin{figure}[t]
  \centering
  \includegraphics[width=0.65\linewidth]{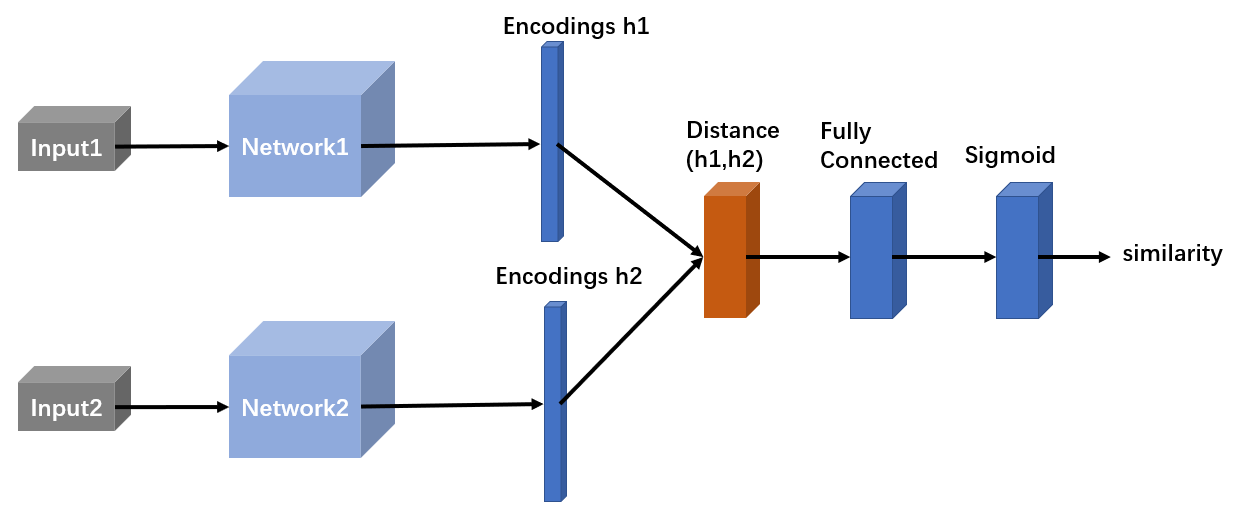}
  \caption{A general architecture of Siamese networks.}
  \label{fig:Siamese network}
\end{figure}

Siamese networks are a type of artificial neural network that measures the distance between the features of two input images to classify them based on their similarity~\cite{chicco2021siamese}. 
Siamese networks learn to measure the distance between the features in two inputs. 
As shown in Fig.~\ref{fig:Siamese network}, the structure of the Siamese network consists of two sub-networks, \emph{Network1} and \emph{Network2}, which share the same weights and biases.
During training, if two inputs are from the same class, the network tries to minimize the distance between them, while if they are from different classes, the network tries to maximize the distance. The output of the network is determined by a Sigmoid function, and based on the distance of Input1 and Input2, the machine learning model can classify them. In exercise detection, Siamese networks can differentiate between exercise and non-exercise data pairs by measuring their distance.

\subsubsection{RNN Networks}\label{subsubsec:rnn}

Recurrent Neural Networks~(RNN)  
are a type of neural network specifically designed to process sequential data, where the current output depends on the previous state. This characteristic makes them powerful for solving problems with temporal dependencies~\cite{chen2021mpcguided,chen2022rfcam}. In the case of physiological signals, consecutive time steps are highly correlated, so RNNs can be useful in revealing hidden exercise information. For this study, Long Short-Term Memory Networks (LSTMs), a specific type of RNN, were used as Network1 and Network2 in Fig.~\ref{fig:Siamese network}.
The general execution of LSTM can be described using the following equations~\cite{munir2021carfi}:
\begin{align}
    i_t &= \sigma(W_{ii}x_t + b_{ii} + W_{hi}h_{t-1} + b_{hi}) \label{equ:lstm1}\\
    f_t &= \sigma(W_{if}x_t + b_{if} + W_{hf}h_{t-1} + b_{hf}) \label{equ:lstm2}\\
    g_t &= tanh(W_{ig}x_t + b_{ig} + W_{hg}h_{t-1} + b_{hg}) \label{equ:lstm3}\\
    o_t &= \sigma(W_{io}x_t + b_{io} + W_{ho}h(t-1) + b_{ho}) \label{equ:lstm4}\\
    c_t &= f_t \odot c_{t-1} + i_t \odot g_t \label{equ:lstm5}\\
    h_t &= o_t \odot tanh(c_t) \label{equ:lstm6}
\end{align}
The LSTM network is capable of storing and retrieving information over long periods of time, thanks to its memory cell (denoted as $c_t$ in the equations). 
Equations~\ref{equ:lstm1} and ~\ref{equ:lstm2} 
compute the input and forget gates, respectively, which determine what information is saved to and forgotten from the memory cell.
Equation~\ref{equ:lstm3} computes the input modulation gate, which regulates the impact of new input data on the memory cell.
Equation~\ref{equ:lstm4} computes the output gate, which determines what information from the memory cell is output as the network's final state. 
The memory cell is updated using Equation~\ref{equ:lstm5}, which combines the input gate and input modulation gate to update the stored information, and the forget gate to selectively delete old information. 
Finally, Equation~\ref{equ:lstm6} computes the output state of the LSTM, which depends on the memory cell and the output gate.


The input to the LSTM consists of sequences of GSR, heart rate, steps taken, and skin temperature data collected by wristbands. 
To preprocess the data, a sliding-window average method is used to mitigate sensor noise, and the input sequences are normalized to the interval $(0, 1)$. 
Data collected during sleeping is eliminated because it introduces bias into the data normalization due to the slower heart rates during sleep.


\subsubsection{Siamese Models for Exercise Detection}

In the Siamese network shown in Fig.~\ref{fig:Siamese network}, two identical LSTM networks are used with two inputs. 
During the training process, two inputs are either from the same class (non-exercise or exercise data) or from different classes (non-exercise data + exercise data). 
At time $t$, the output vectors of the two LSTM networks are denoted as $v_{1,t}$ and $v_{2,t}$, respectively. 
The difference between these two vectors is calculated as ${v}_t = ||v_{1,t} - v_{2,t}||_2$.

Next, $v_t$ is fed into a fully connected layer followed by a Sigmoid function. 
The output of the Siamese network falls in the range of $(0, 1)$. 
After training, the Siamese network learns the similarities and differences between the training samples. 
The optimizer used is the stochastic gradient descent (SGD) algorithm, with the loss function being binary cross-entropy.
The fundamental idea behind our machine learning model is to learn weights and biases that enable it to minimize the distance between samples from the same class (e.g., non-exercise data) and maximize the distance between samples from different classes (i.e., non-exercise data and exercise data). In each epoch, the input samples are labeled as either from the same class or from different classes, so that the model can learn to classify the input data.

%% file: glucose_model.tex
\subsection{Data-driven Physiological Models}

\begin{figure*}[t]
  \centering
  \includegraphics[width=0.6\linewidth]{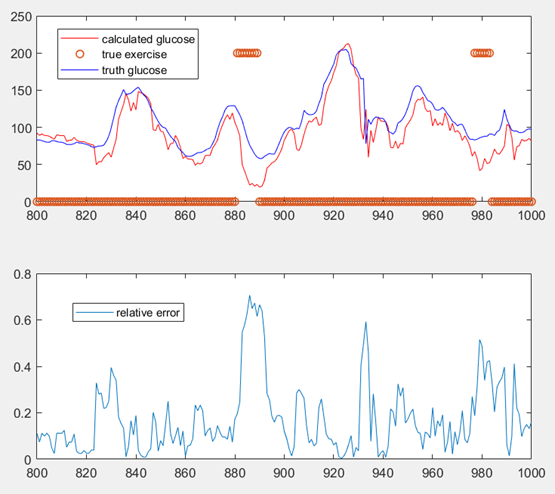}
  \caption{Relative errors between model-based predicted glucose levels and CGM sensor readings are indicators of exercises.}
  \label{fig:error}
\end{figure*}

While Siamese networks may be effective in detecting exercise using various sensors, they may not always be reliable. For instance, the step counts of a patient riding a stationary bike may remain unchanged, and high room temperatures may cause an increase in skin temperature readings, even if the patient is not exercising~\cite{Neves2015DifferentRO,gisolfi1984temperature,torii1992fall}. 
This implies that factors other than exercise may have effects similar to those of exercise on sensor readings. Thus, to accurately detect exercise, it is necessary to consider BG readings from CGM, and we propose a model that describes the effect of exercise on the BG readings of T1DM patients. 
Based on previous work~\cite{breton2008physical,roy2007dynamic}, we propose a new model that takes the effects of exercise and meal intake on glucose into account, as described below. The following equations describe the remote insulin distribution and plasma insulin distribution from a biological perspective:

\begin{align}
\dot{G}(t) &= -p_{1}\cdot (G(t)-G_{b}) - p_{4}X(t)G(t) - p_{6}y(t)G(t) +  u_{2}(t)/V_{g} \label{eq:G(t)}\\
\dot{X}(t) &= -p_{2}X(t) + p_{3}\cdot (I(t) - I_{b})\label{eq:X(t)}\\
\dot{I}(t) &= -m\cdot (I(t) - I_{b}) + p_{5}u_{1}(t) \label{eq:I(t)} \\
\dot{y}(t) &= -1/\tau_{HR} \cdot y(t) + 1/\tau_{HR} \cdot (HR(t) - HR_{b}) \label{eq:y(t)}
\end{align}
In these equations, $I(t)$, $X(t)$, $G(t)$, and $HR(t)$ represent plasma insulin, remote insulin, plasma glucose, and heart rate, respectively.
Variables $u1(t)$ and $u2(t)$  represent endogenous insulin infusion and dietary absorption, respectively. 
Equation~(\ref{eq:G(t)})  quantifies the effect of exercise and associates this effect with BG levels.
The factor $HR_{b}$ represents the basal heart rate, and we set $HR_{b} = 60$ bps, as in Dalla Man et al.~\cite{dalla2009physical}. The factor $I_{b}$ represents the basal plasma insulin~\cite{roy2007dynamic}. 
We set $1/\tau_{HR}= 5$, as in its original definition~\cite{breton2008physical}.
The parameter $G_{b}$ represents the basal BG level or fasting BG level, while $V_{g}$ represents the glucose distribution.   
Table~\ref{table:Parameters known} displays the known parameters in our model, while $p_1$, $p_4$, and $p_6$ are the parameters that need estimation. The left sides of Equations~(\ref{eq:G(t)})--(\ref{eq:y(t)}) are in derivative formations, so we discretize them before applying the minimal mean square error estimation technique to estimate the parameters.

are in derivative formations, so we discretize them before applying the minimal mean square error estimation technique to estimate the parameters. 
We use Equation~(\ref{eq:G(t)}), to predict $G(t+1)$ using $G(t)$.
When exercise occurs, the predicted value may be inaccurate. To evaluate the inaccuracy, we calculate the relative error as follows.
\begin{align}
    error(t) = \frac{|G(t+1)-G_{gt}(t+1)|}{G_{gt}(t+1)}
\end{align}
where $G_{gt}(t)$ is the ground truth of $G(t)$ obtained from CGM.
As shown in Fig.~\ref{fig:error}, the error values are relatively high during exercise. Besides relative errors, we also consider heart rate values and absolute glucose levels. To detect exercise using the model-based detector, we design Algorithm~\ref{alg:ensemble}. 
In this algorithm, there are three constraints on heart rate, absolute glucose levels, and glucose relative error. When all constraints are met for at least $f_{th}$ consecutive time steps, the model-based detector determines that an exercise is happening 

\begin{algorithm}[t]
  \caption{Exercise Detection with Model-based Detector}
   \label{alg:ensemble}
    \KwIn{Plasma glucose levels $G(t)$, heart rate $HR(t)$, experiment time $L$, threshold $f_{th}$
    }
    Initialize $flag=0$. \label{alg:init}\\
    \For{$t = 1$ to $L$ }{
    \If{$HR(t)\geq 85$ and $|G(t)-G(t-1)|\leq 0$ and $error(t)\geq 0.35$}{
    $flag ++$ \\
    }
    \Else{$flag=0$}
    \If{$flag\geq f_{th}$}{
    Announce an exercise event;\\
    }
    } 
\end{algorithm}



\begin{table}[t]
    \centering
    \caption{Known parameters in our insulin-glucose physiological model}
    \label{table:Parameters known}
\begin{tabular}{|p{19pt}|p{40pt}|p{45pt}|p{110pt}|p{18pt}|p{12pt}|p{10pt}|}
\hline
 $p_{2}$ & $p_{3}$ & $p_{5}$ & $G_{b}$, $I_{b}$ & $m$ & $\tau_{HR}$ & $V_{g}$ \\
 \hline
0.037 & $1.2\times10^{-5}$ & $5.68\times10^{-4}$ & Calculate for each patient & 0.142 & 0.2 & 117\\
 \hline
\end{tabular}
\end{table}

%% file: combine_both_models.tex
\subsection{Ensemble Learning-based Approach}

Ensemble learning is a popular approach to improve the accuracy and robustness of machine learning models~\cite{sagi2018ensemble}. By combining multiple models, it is possible to reduce individual model biases and increase the overall accuracy of the ensemble. In this context, we propose an ensemble learning-based approach that combines the strengths of both the Siamese network and the data-driven physiological model to achieve faster and more accurate exercise detection.
The Siamese network is used as the first detector of exercise events, given its low latency and real-time detection capabilities. However, it has a high false detection rate due to its limited ability to differentiate between exercise and other activities that might lead to similar movement patterns. On the other hand, the model-based detector has a longer detection delay but provides high accuracy in detecting exercise events. Therefore, we aim to combine the two methods to overcome their individual limitations and improve the overall performance of the exercise detection system.

\begin{figure*}[t]
  \centering
  \includegraphics[scale=0.5]{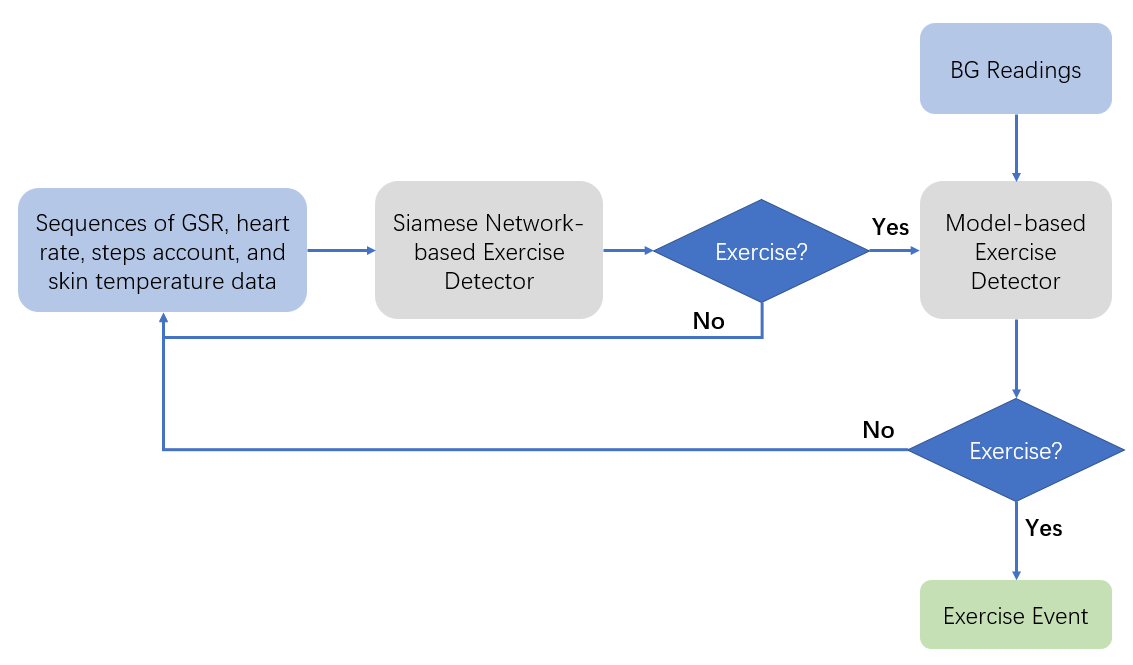}
  \caption{Ensemble learning using the Siamese network-based and model-based exercise detectors.}
  \label{fig:ensemble}
\end{figure*}

In our approach, the Siamese network is used as the first stage of detection. Once an exercise event is detected by the Siamese network, the model-based detector is activated to perform a more accurate detection. This two-stage detection approach allows us to achieve high accuracy in detecting exercise events while maintaining low detection latency. The proposed ensemble learning-based approach is shown in Fig.~\ref{fig:ensemble}, where the output of the Siamese network is used as input to the model-based detector, which then generates the final detection result.
The proposed ensemble learning-based approach is a promising direction for improving the performance of exercise detection systems, as it combines the strengths of different models to achieve better accuracy and robustness.

%% file: experiment_setup.tex
\section{Experimental Evaluation}\label{sec:eval}

In this section, we will provide details on the dataset and methodology used to evaluate the proposed approaches.

\subsection{The OhioT1DM Dataset}

The OhioT1DM Dataset is a valuable resource for research in blood glucose level prediction, as it provides eight weeks of data for 12 individuals diagnosed with type 1 diabetes. The dataset includes a range of data fields, such as glucose level, finger stick, basal, temporary basal, bolus, meal, sleep, work, stress, hypo event, illness, exercise, heart rate, GSR, skin temperature, air temperature, steps, basis sleep, and acceleration. It is important to note that some of the data fields are collected through self-reported methods, and the data is collected using various types of fitness bands, resulting in differing levels of data for each patient. To address the heterogeneity in the data, innovative approaches are needed to accurately predict blood glucose levels for type 1 diabetes patients. The dataset is available in XML format and is divided into training and testing sets, making it an accessible resource for researchers in the field. Table~\ref{table:patient info} provides a detailed breakdown of the data fields collected from each patient, further highlighting the richness of the dataset.

\begin{table}[t]
    \caption{Data fields of different patients in the OhioT1DM dataset}
    \label{table:patient info}
\begin{tabular}{|p{60pt}|p{290pt}|p{60pt}|}
 \hline
  Patients & Data fields & Fitness Band\\
 \hline
 559, 563, 570, 575, 588, 591 & Glucose, Finger Stick, Basal, Temp\_basal, Bolus, Meal, Sleep, Work, Stressor, Hypo\_event, Illness, Exercise, Heart Rate, GSR, Skin Temperature, Air Temperature, Step Counts, Basis\_sleep & Basis      \\
 \hline
 540, 544, 552, 567, 584, 596 & Glucose, Finger Stick, Basal, Temp\_basal, Bolus, Meal, Sleep, Work, Stressor, Hypo\_event, Illness, Exercise, GSR, Skin Temperature, Acceleration & Empatica  \\
\hline
\end{tabular}
\end{table}

\subsection{Experimental Setup}
In this study, the OhioT1DM Dataset was used for both training and testing the proposed model. To process the data, essential fields such as GSR, insulin, meal, exercise, step count, skin temperature, heart rate, and glucose readings were retrieved. These fields were sampled every 5 minutes in the OhioT1DM Dataset, and data collected during sleep was discarded, as explained in Section~\ref{subsubsec:rnn}. However, the data sampling intervals in the OhioT1DM Dataset may vary, which makes the alignment of samples a challenging task. Additionally, missing data are common in realistic datasets. To address these issues, a time difference of less than 5 minutes was considered while aligning data in two samples. Furthermore, a Savitzky-Golay filter was applied to the heart rate and GSR values to remove noise, and missing data were filled in. To avoid mislabeling of exercise sessions by patients, the ground truth exercise sessions were smoothed. The data was then divided into two groups, signals during exercise and signals during non-exercise, to form input pairs for the Siamese network-based detector training.

%% file: experiment_result.tex
\subsection{Experimental Results}

In this section, we evaluate the performance of our exercise detector using ensemble learning and compare it with four baseline detectors. We assess the performance of the detectors by comparing their confusion matrices to classify exercise and non-exercise events. Additionally, we investigate the impact of system hyper-parameters on performance.

\subsubsection{Performance Evaluation}

In this section, we evaluate the performance of our Ensemble Learning-based exercise detector and compare it with four baseline detectors, namely: 1) our Siamese network-based detector; 2) our model-based detector; 3) a detector based on energy expenditure~\cite{dasanayake2015early}; and 4) a detector based on heart rate and accelerometer data~\cite{zakeri2008application}.

\begin{figure}[t]
    \centering
    \begin{minipage}{.45\linewidth}
    \centering
    \includegraphics[width =\linewidth]{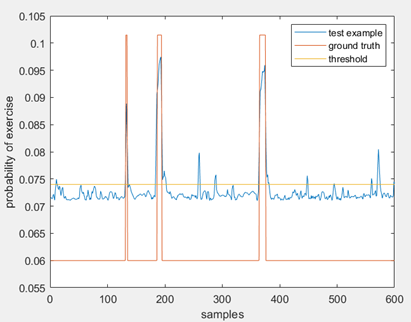}
    \caption{Exercise detection using the Siamese network.}
    \label{fig:siamese}
    \end{minipage}\hfill
    \begin{minipage}{.45\linewidth}
    \centering
    \includegraphics[width = .95\linewidth]{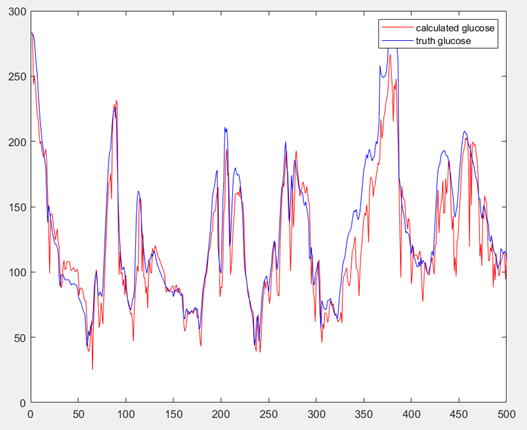}
    \caption{Predicted trajectory using our glucose model.}
    \label{fig:model}
    \end{minipage}
\end{figure}

For the Siamese network-based detector, one input serves as an "anchor", which is the mean value of 100 non-exercise samples selected randomly from the training set. The Siamese network-based detector calculates the difference/similarity between the test samples and the anchor to determine whether the test samples indicate an exercise event. Fig.~\ref{fig:siamese} shows the output of the Siamese network-based detector in blue and the ground truth of exercise samples in red. A threshold (yellow line) is set to determine if an exercise event is detected. As shown in the figure, the output probability is much higher during exercise events, indicating that the detector performs well in detecting exercise events.
For the model-based detector, we use Algorithm~\ref{alg:ensemble} to detect exercise events. Fig.~\ref{fig:model} shows that our data-driven physiological model predicts BG (red line) that closely matches the ground truth (blue line).

\begin{figure}[t]
    \centering
    \subfloat[]{
    \includegraphics[width = .24\linewidth]{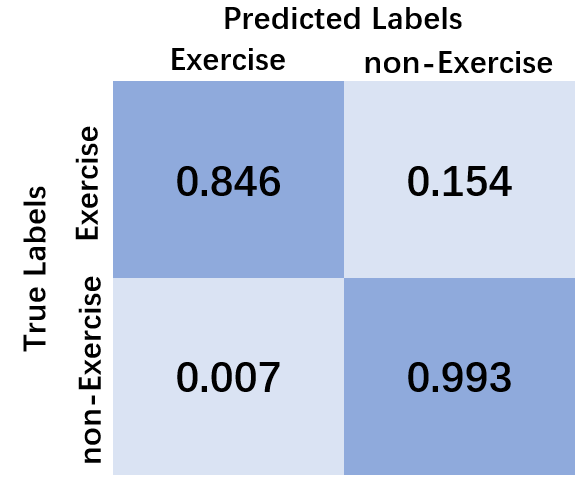}
    \label{fig:matrix1}
    }
    \subfloat[]{
    \includegraphics[width = .24\linewidth]{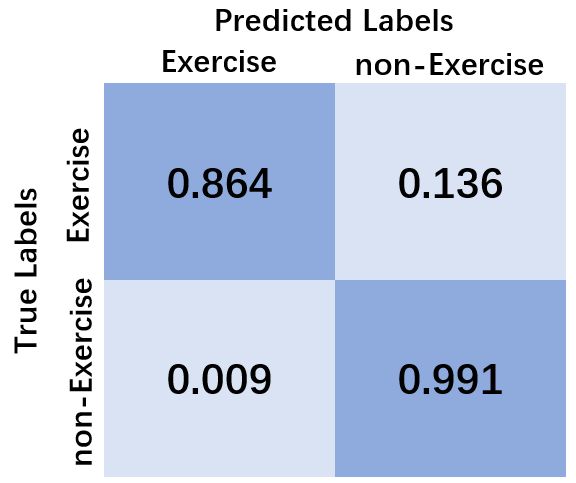}
    \label{fig:matrix2}
    }
\subfloat[]{
    \includegraphics[width = .24\linewidth]{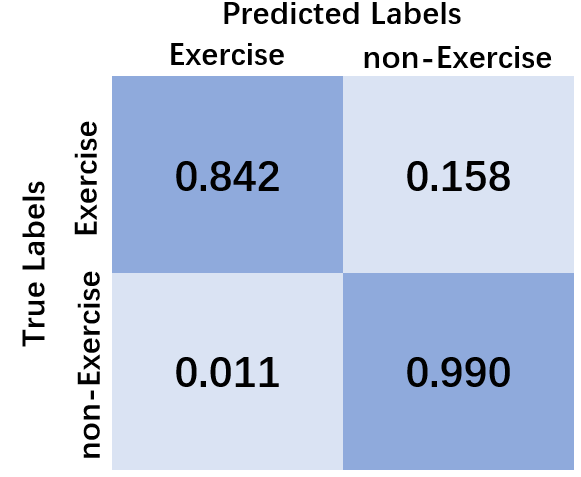}
    \label{fig:matrix3}
    }
    \subfloat[]{
    \includegraphics[width = .24\linewidth]{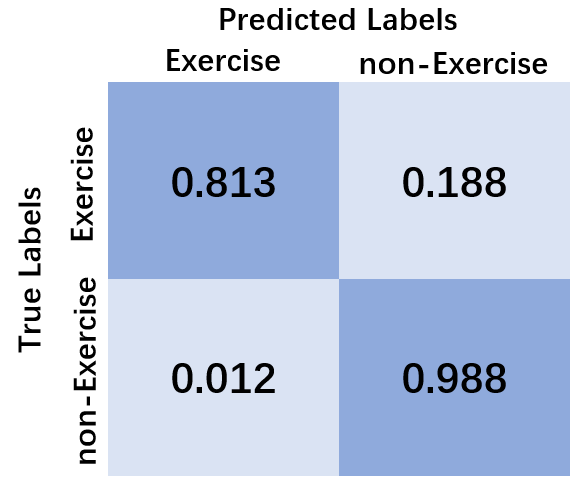}
    \label{fig:matrix4}
    }
    \caption{Exercise v.s. Non-Exercise confusion matrices when threshold $f_{th}$ of our exercise detector is set to be 2, 3, 4, and 5.}
    \label{fig:matrix_signal}
\end{figure}

\begin{figure}[t]
    \centering
    \subfloat[]{
    \includegraphics[width = .24\linewidth]{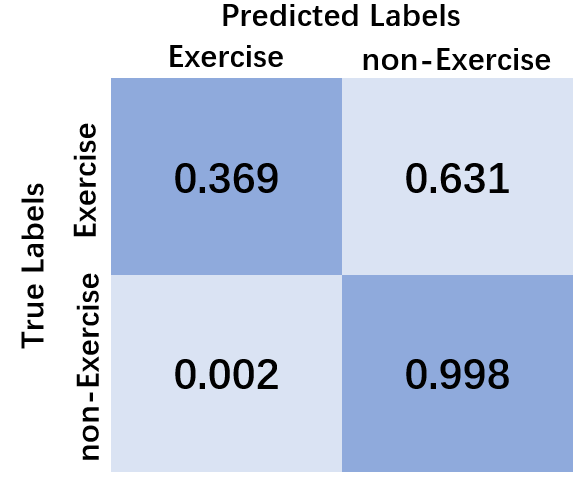}
    \label{fig:matrix_siamese}
    }
    \subfloat[]{
    \includegraphics[width = .24\linewidth]{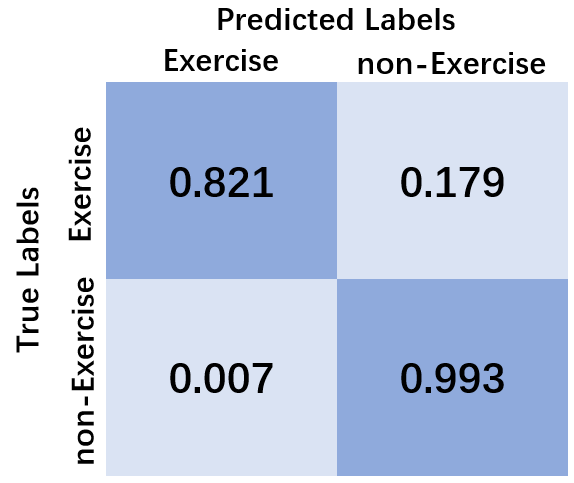}
    \label{fig:matrix_model}
    }
    \subfloat[]{
    \includegraphics[width = .24\linewidth]{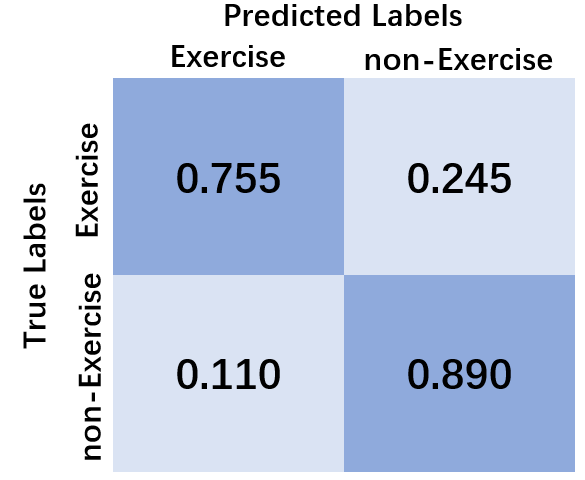}
    \label{fig:matrix_base1}
    }
    \subfloat[]{
    \includegraphics[width = .24\linewidth]{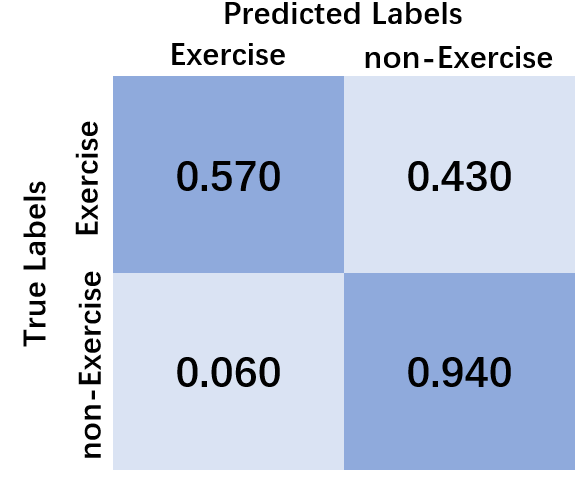}
    \label{fig:matrix_base2}
    }
    \caption{Exercise v.s. Non-Exercise confusion matrices using baseline detectors. (a)~Our Siamese network-based detector. (b)~Our model-based detector. (c)~An energy expenditure-based detector. (d)~A detector based on heart rate and acceleration data.}
    \label{fig:matrix_baseline}
\end{figure}

We calculate the confusion matrices of our approach and four baselines, as shown in Tables~\ref{fig:matrix_signal} and~\ref{fig:matrix_baseline}.
In Fig.~\ref{fig:matrix2}, our approach achieves a true positive rate of $86.4\%$ and a true negative rate of $99.1\%$, outperforming all baselines shown in Fig.~\ref{fig:matrix_baseline} in accurately differentiating between exercise and non-exercise events.

\subsubsection{Effect of Hyper-parameters}

In this section, We evaluate how the detection accuracy of our exercise detector is affected by the hyper-parameter $f_{th}$ (see Algorithm~\ref{alg:ensemble}).
As shown in Figs.~\ref{fig:matrix1} and~\ref{fig:matrix2},  we observe that the true positive rate and the true negative rate improve with a larger $f_{th}$ value. This is because exercise has a long-lasting effect on glucose, and longer signal sequences can better capture the presence of exercise. 
However, we also find that the true positive rate and the true negative rate decrease with larger $f_{th}$, as illustrated in Figs.~\ref{fig:matrix3} and~\ref{fig:matrix4}.
A deeper investigation reveals that patients often have meals after exercise, which can cause glucose levels to rise. Therefore, the detection of meal events can benefit the performance of exercise detection when BG data is available.

%% file: future_work.tex
\section{Future work}

During the course of our experiments, we observed that meal intake can have a significant impact on BG data and hence, exercise detection. Many patients consume meals after exercise, and some even have snacks before exercising. In future work, we plan to investigate the impact of meal intake on exercise detection for T1DM patients. One potential approach is to incorporate model-based meal detection using BG readings~\cite{chen2019committed} into our exercise detector. Another possible avenue is to explore eating detection using wearable acceleration sensors~\cite{ye2015automatic}. By taking into account meal intake, we hope to improve the accuracy and robustness of our exercise detector and enable it to better serve the needs of T1DM patients.


%% file: Conclusion.tex
\section{Conclusion}

In this paper, we have presented an ensemble learning approach for detecting exercise events in T1DM patients that combines a Siamese network-based detector and a model-based detector. The proposed approach was evaluated on the OhioT1DM dataset, and achieved a true positive rate of $86.4\%$ and a true negative rate of $99.1\%$, outperforming state-of-the-art solutions. Our results suggest that combining the strengths of different detectors can lead to improved performance in exercise detection. Future work will focus on incorporating meal intake detection into the proposed approach to further enhance its accuracy in real-world scenarios.
